\documentclass[conference,a4paper]{IEEEtran}

\usepackage{amssymb, amsmath,latexsym}
\usepackage[dvips]{graphicx}

\newtheorem{thm} {Theorem}
\newtheorem{lem} {Lemma}

\def\R{{\mathbb{R}}}
\def\Z{{\mathbb{Z}}}
\def\N{{\mathbb{N}}}

\def\dim{{{\rm dim}}}

\def\R{\mathbb{R}}

\begin{document}

\sloppy

\title{Second Order Asymptotics for Random Number Generation }


\author{
   \IEEEauthorblockN{
      Wataru Kumagai\IEEEauthorrefmark{1,}\IEEEauthorrefmark{2} and
     Masahito Hayashi\IEEEauthorrefmark{2,}\IEEEauthorrefmark{3},
}
   \IEEEauthorblockA{
     \IEEEauthorrefmark{1}Graduate School of Information Sciences, Tohoku University, Japan\\ 
     Email: wkumagai1001@gmail.com}
   \IEEEauthorblockA{
     \IEEEauthorrefmark{2}Graduate School of Mathematics, Nagoya University, Japan\\
   Email: masahito@math.nagoya-u.ac.jp}
   \IEEEauthorblockA{
     \IEEEauthorrefmark{3}Centre for Quantum Technologies, National University of Singapore, Singapore\\
     }
 }



\maketitle


\begin{abstract}
We treat a random number generation from an i.i.d. probability distribution of $P$ to that of $Q$.
When $Q$ or $P$ is a uniform distribution, the problems have been well-known as the uniform random number generation and the resolvability problem respectively, and analyzed not only in the context of the first order asymptotic theory but also that in the second asymptotic theory.
On the other hand, when both $P$ and $Q$ are not a uniform distribution, the second order asymptotics has not been treated. 
In this paper, we focus on the second order asymptotics of a random number generation  for arbitrary probability distributions $P$ and $Q$ on a finite set.
In particular, we derive the optimal second order generation rate under an arbitrary permissible confidence coefficient.
\end{abstract}

\section{Introduction}

The random number generation is one of the most basic problems in the information theory.
The purpose of the random number generation is to approximate a sequence of target probability distributions $Q_n$ by transforming a sequence of another probability distributions $P_n$. 
When $Q_n$ or $P_n$ is a uniform distribution, each problem corresponds to the uniform random number generation problem or the resolvability problem respectively, and has been well studied.
For example, when $P_n$ and $Q_n$ are the i.i.d. probability distributions $P^n$ and $U_2^{an}$ where $U_2$ is the uniform distribution with the support size $2$, the optimal first order generation rate $a$ from $P^n$ is the entropy $H(P)$ under the condition that the error goes to $0$.
Those problems are analyzed not only in the context of the first order asymptotic theory but also that in the second asymptotic theory.
In the most general case, it is known that the first and the second order optimal rates in those problems can be described by the information spectrum methods \cite{Han, HN, Hay1}.
In particular, for transformation between $P^n$ and $U_2^{an+b\sqrt{n}}$, the results in the information spectrum gives optimal rates $a$ and $b$.
On the other hand, when both $P_n$ and $Q_n$ are not a uniform, the problem has not been treated sufficiently. 
In this paper, we do not restrict both $P_n$ and $Q_n$ to a uniform distribution and focus on the second order asymptotics of a random number generation for arbitrary i.i.d. probability distributions on a finite set.
In particular, we derive the optimal second order generation rate under an arbitrary permissible confident coefficient.

In this paper, we utilize the notion of the majorization. 
It is a pre-order between two probability distributions which can be defined on different finite sets. 
If a probability distribution $P_n$ is transformed to $W_n(P_n)$ by a deterministic transformation $W_n$, the transformed probability distribution $W_n(P_n)$ "majorizes" the original probability distribution $P_n$.
In other words, $W_n(P_n)$ is larger than $P_n$ in the sense of the majorization relation.
Therefore, when we want to approximate a target probability distribution $Q_n$ from an original probability distribution $P_n$, for an arbitrary deterministic transformation $W_n$, there is a probability distribution $P'_n$ which majorizes $P_n$ and is close to $Q_n$ than $W_n(P_n)$.
Thus, the performance of the optimization under the majorization condition gives a bound of that under deterministic transformations. 
The majorization is used in a transformation theory of quantum entangled states and corresponds to a operation called LOCC in the quantum information theory \cite{Nie, JP}.
Our results can be extended to the quantum settings but we do not mention it in this paper.

The paper is organized as follows.
In section II, we introduce a notion of majorization, and consider approximation problems under a majorization condition and by a deterministic transformation.
In section III, we treat the first order asymptotics for the approximation problem and derive the first order optimal rate under the i.i.d. setting. 
In section IV, we review the existing studies about the second order
asymptotics of the approximation problem when the source distribution or the target distribution is a uniform distribution. 
In section V, we treat the second order asymptotics of the approximation problem when both
the source distribution and the target distribution are not a uniform distribution.
We note that the results itself in section V do not contain that in section IV.
But, in the end of section V,  we show that the result in Section IV can be regarded as the limit case of the sesult in section V. 
In section VI, we state the conclusion of the paper.

\section{One-Shot Formulation}~

In this section, we introduce some notation and definition, and formulate our problem. 
For a probability distribution $P$ on finite set $\mathcal{X}$ and a map $W:\mathcal{X}\to\mathcal{Y}$, the probability distribution $W(P)$ on $\mathcal{Y}$ is defined by $W(P)(y):=\sum_{x\in W^{-1}(x')}P(x)$.
We introduce a value $F$ called the Bhattacharyya coefficient or the fidelity between probability distributions over the same discrete set $\mathcal{Y}$ as
\begin{eqnarray}
F(Q,Q'):=\sum_{y\in\mathcal{Y}}\sqrt{Q(y)}\sqrt{Q'(y)}.
\end{eqnarray}
This value $F$ represents how close two probability distributions are and relates to the Hellinger distance $d_H$ as $d_H(\cdot,\cdot)=\sqrt{1-F(\cdot,\cdot)}$.
Then our main purpose is to analyze the following value.  
\begin{eqnarray}
L^D(P,Q|\nu)
:=\max\{L|F(W(P),Q^L)\ge\nu, W:\mathcal{X}\to\mathcal{Y}^L\}.\hspace{-1em}
\end{eqnarray}
This means the maximal number $L$ of $Q^L$ which can be approximated from $P$ under a confidence coefficient $0<\nu<1$.
When we define the maximal fidelity $F^D$ from $P$ on $\mathcal{X}$ to $Q$ on $\mathcal{Y}$ by
\begin{eqnarray}
F^D(P\to Q):=\max\{F(W(P),Q)|W:\mathcal{X}\to\mathcal{Y}\},
\end{eqnarray}
Then $L^D$ is rewritten as
\begin{eqnarray}
L^D(P,Q|\nu)
=\max\{L|F^D(P\to Q^L)\ge\nu\}.
\end{eqnarray}

Next, we will introduce the notion of the majorization to evaluate $F^D$.
For a probability distribution $P$ on a finite set, let $P^{\downarrow}$ be a sequence $\{P^{\downarrow}_i\}_{i=1}^{\infty}$ where $P^{\downarrow}_i$ is the element of $\{P(x)\}_{x\in\mathcal{X}}$ sorted in decreasing order for $1\le i\le|\mathcal{X}|$ and $P^{\downarrow}_i$ is $0$ for $|\mathcal{X}|< i$.
We set as $C_{P}(l):=\sum_{i=1}^lP^{\downarrow}_i$.
When probability distributions $P$ and $Q$ satisfy $C_{P^{\downarrow}}(l)\le C_{Q^{\downarrow}}(l)$ for any $l\in\N$, it is said that $P$ is majorized by $Q$ and written as $P\prec Q$.
Here, note that the sets where $P$ and $Q$ are defined do not necessarily coincide with each other.
The majorization relation is a pre-order on a set of probability distributions in which each distribution is defined on a finite set\cite{MO}. 
We introduce the maximal fidelity under the majorization condition.
\begin{eqnarray}
F^M(P\to Q):=\max\{F(P',Q)|P\prec P'~on~\mathcal{Y}\}
\end{eqnarray}
where $P$ and $Q$ are probability distribution on $\mathcal{X}$ and $\mathcal{Y}$, respectively.
Since $P\prec W(P)$ for a map $W:\mathcal{X}\to\mathcal{Y}$, 
\begin{eqnarray}
F^M(P\to Q)\ge F^D(P\to Q)\label{fidelity ineq}
\end{eqnarray}
holds.
In many case, $F^M$ is easily treatable than $F^D$.
In particular, the value of $F^M$ can be explicitly described in one-shot situation \cite{VJN}.

\section{First Order Asymptotics}~

We proceed to the asymptotics of random number generation.
Here, we assume the form $P^n$ and $Q^{an}$ as $P$ and $Q$ in the last section. 
Then, we call $a>0$ the first order rate.
In this section, we derive the dependency of $F^D(P^{n} \to Q^{an})$ for the first order rate $a$ when $n$ goes to the infinity.
The following theorem say that the threshold value of the first order rate is the ratio of the entropy of $P$ and $Q$.
\begin{thm}\label{first order error}
\begin{eqnarray}
{\lim}F^D(P^{n} \to Q^{an})
&=&{\lim}F^M(P^{n} \to Q^{an})\nonumber\\
&=&\left\{
\begin{array}{ll}
1 & \mathrm{if}~a<\frac{H(P)}{H(Q)}\\
0 & \mathrm{if}~a>\frac{H(P)}{H(Q)}.
\end{array}
\right.
\end{eqnarray}
\end{thm}
The limit of the maximal fidelity is obtained by Theorem \ref{first order error} when the first order rate is not $H(P)/H(Q)$.
We rewrite $L^D(P^n,Q|\nu)$ as $L^D_n(P,Q|\nu)$. 
Then, we can not obtain even the first order asymptotic expansion of $L^D_n(P,Q|\nu)$ for $0<\nu\le 1$ from Theorem \ref{first order error} because the case then $a=H(P)/H(Q)$ is not treated in Theorem \ref{first order error}.
In other words, we can not derive the performance of $L^D_n(P,Q|\nu)$ under a confidence coefficient $0<\nu\le 1$ when $n$ is large.
Therefore, we fix the first order rate $a$ to $H(P)/H(Q)$ and analyze the second order rate in the next section.

\section{Second Order Asymptotics for Uniform Distribution}~

We focus on the approximation from $P^n$ to $Q^{an+b\sqrt{n}}$ and assume that the first order rate $a$ is $H(P)/H(Q)$.
We treat the following problems.\\
(I)~For a fixed second order asymptotic rate $b\in\R$ of $Q^{H(P)/H(Q)n+b\sqrt{n}}$, we derive the limit value of the maximal fidelity $\lim F^D(P^{n} \to Q^{H(P)/H(Q)n+b\sqrt{n}})$.\\
(II)~For a fixed confidence coefficient $0<\nu<1$, we derive the second order asymptotic expansion of $L^D_n(P,Q|\nu)$.\\
These problems are essentially equivalent. 
To derive the second order asymptotic expansion under a confidence coefficient $0<\nu<1$, we focus on the following value which means the optimal second order rate.
\begin{eqnarray}
&&\hspace{-2em}R^D_2(P,Q|\nu)\nonumber\\
&&\hspace{-2em}:=\sup\left\{b\in\R\Big|\mathrm{lim} F^D(P^n\to Q^{\frac{H(P)}{H(Q)}n+b\sqrt{n}})\ge\nu\right\}.
\end{eqnarray}
If the above value is finite, the second order asymptotic expansion of $L^D_n(P,Q|\nu)$ is represented as 
\begin{eqnarray}
&&\hspace{-2em}L^D_n(P,Q|\nu)\nonumber\\
&&\hspace{-2em}=(H(P)/H(Q))n+R^D_2(P,Q|\nu)\sqrt{n}+o(\sqrt{n}).\label{asym.expansion2}
\end{eqnarray}

We review the results for the uniform random number generation and the resolvability in \cite{HN, Hay1}.
Let $U_2$ be the uniform distribution with the support size $2$. 
Then the limit value of the maximal fidelity is represented as follows for a fixed second order rate $b\in\R$.
\begin{thm}\label{2-order error}
Let $P$ be any probability distribution on a finite set except for a uniform distribution. Then
\begin{eqnarray}
&& \lim F^D(P^{ n} \to U_2^{ H(P)n+b\sqrt{n}})\nonumber\\
&&\hspace{-1.5em}=\lim F^M(P^{ n} \to U_2^{ H(P)n+b\sqrt{n}})
=\sqrt{1-G\left(\frac{b}{\sqrt{V(P)}}\right)},  \nonumber\label{2-order1}
\end{eqnarray}
where $H(P)$ is the entropy of $P$ and 
\begin{eqnarray}
&&V(P):=\displaystyle\sum_{x\in\mathcal{X}}P(x)(-\mathrm{log}P(x)-H(P))^2.
\end{eqnarray}
\end{thm}

From Theorem \ref{2-order error}, it turned out that the limit of the maximal fidelity depend on the second order rate $b$ when $a=H(P)/H(Q)$.
Note that the limit value does not depend on the second order rate $b$ when $a\ne H(P)/H(Q)$ is given by Theorem \ref{first order error}.
We emphasize that the lower order term does not affect the limit value if $an+b\sqrt{n}$ has lower order term as $an+b\sqrt{n}+o(\sqrt{n})$ (e.g. $o(\sqrt{n})=\log n$). 
Hence, when we want to analyze the maximum fidelity, we only have to treat the first and second order rate and do not need the third order asymptotics. 
For a fixed confidence coefficient $0<\nu<1$, the second order asymptotic expansion is represented as follows. 
\begin{thm}\cite{Hay1}\label{2-order rate}
Let $P$ be any probability distribution on a finite set except for a uniform distribution. Then the second order asymptotic expansions in (\ref{asym.expansion2}) are described as follows. 
\begin{eqnarray}\label{exp.con}
&&\hspace{-2em} L^D_n(P,U_2|\nu)\nonumber\\
&&\hspace{-2em}=H(P)n-\sqrt{V(P)}G^{-1}(\nu^2)\sqrt{n}+o(\sqrt{n}),
\end{eqnarray} 
where $G$ is the cumulative distribution function of the standard normal distribution. 
Here, 
\begin{eqnarray}
&R^D_2(P,U_2|\nu)=-\sqrt{V(P)}G^{-1}(\nu^2).&\nonumber
\end{eqnarray}
\end{thm}

Next we consider the approximation from the uniform distribution to $P$.
The limit value of the maximal fidelity is represented as follows for a fixed second order rate $b\in\R$.
\begin{thm}\label{2-order error2}
Let $P$ be any probability distribution on a finite set except for a uniform distribution. Then
\begin{eqnarray}
&&\lim F^D(U_2^{n} \to P^{ H(P)^{-1}n+b\sqrt{n}})\nonumber\\
&=&\lim F^M(U_2^{n} \to P^{ H(P)^{-1}n+b\sqrt{n}})\\
&=&\sqrt{G\left(\frac{-H(P)^{\frac{3}{2}}b}{\sqrt{V(P)}}\right)}. \label{2-order2}
\end{eqnarray}
\end{thm}
For a fixed confidence coefficient $0<\nu<1$, the second order asymptotic expansion is represented as follows.
\begin{thm}\cite{HN}\label{2-order rate2}
Let $P$ be any probability distribution on a finite set except for a uniform distribution. Then the second order asymptotic expansions in (\ref{asym.expansion2}) are described as follows. 
\begin{eqnarray}\label{exp.dil}
&&\hspace{-2em} L^D_n(U_2,P|\nu)\nonumber\\
&&\hspace{-2em} =H(P)^{-1}n-\sqrt{\frac{V(P)}{H(P)^3}}G^{-1}(\nu^2)\sqrt{n}+o(\sqrt{n}).
\end{eqnarray}
\end{thm}
Here, 
\begin{eqnarray}
&R^D_2(U_2,P|\nu)
=-\sqrt{\frac{V(P)}{H(P)^3}}G^{-1}(\nu^2).&\nonumber
\end{eqnarray}

\section{Second Order Asymptotics for Non Uniform Distribution}~

In this section, we treat non-uniform distribution cases.
We note that the results itself in this section do not contain that in the section IV because we use the property that both $V(P)$ and $V(Q)$ are not $0$, which is equivalent to that both $P$ and $Q$ are not uniform distributions.
But, as is shown in the end of this section, the result in section IV can be regarded as the limit case of the result in section V. 
We define some notations for non-uniform probability distributions $P,Q$ and a constant $b\in\R$. Theorems which appear later are represented by those symbols.
\begin{eqnarray}
&&\hspace{-1em}N_{P}:=N(0,V(P)), \\
&&\hspace{-1em}N_{P,Q,b}:=N\left(H(Q)b,\frac{H(P)}{H(Q)}V(Q)\right),
\end{eqnarray}
\begin{eqnarray}
&&\hspace{-1em}G_{P}(x):=G\left(\frac{x }{\sqrt{V(P)}}\right),\\
&&\hspace{-1em}G_{P,Q,b}(x):=G\left(\sqrt{\frac{H(Q)}{H(P)V(Q)}}(x-H(Q)b)\right),
\end{eqnarray}
\begin{eqnarray}
&&\hspace{-1em}I_{P,Q,b}(x):=\sqrt{\frac{2\sqrt{C_{P,Q}}}{1+C_{P,Q}}}e^{-\frac{(H(Q)b)^2}{4V(P)(!+C_{P,Q})}}\nonumber\\
&&~~~~~\times G\left(\sqrt{\frac{1+C_{P,Q}}{2V(P)C_{P,Q}}}\left(x-\frac{H(Q)b}{1+C_{P,Q}}\right)\right),\\
&&\hspace{-1em}I_{P,Q,b}(\infty):=\sqrt{\frac{2\sqrt{C_{P,Q}}}{1+C_{P,Q}}}e^{-\frac{(H(Q)b)^2}{4V(P)(!+C_{P,Q})}}
\end{eqnarray}
where $N(\mu,v)$ is the normal distribution with the mean $\mu$ and the variance $v$, and $C_{P,Q}:=\frac{H(P)}{V(P)}\left(\frac{H(Q)}{V(Q)}\right)^{-1}$.
Note that $G_{P},G_{P,Q,b}$ means the cumulative distribution functions of $N_{P},N_{P,Q,b}$ and
\begin{eqnarray}
&I_{P,Q,b}(x)=\displaystyle\int_{-\infty}^{x}\sqrt{N_P(t)}\sqrt{N_{P,Q,x}(t)}dt,&\\
&I_{P,Q,b}(\infty)=\displaystyle\int_{-\infty}^{\infty}\sqrt{N_P(t)}\sqrt{N_{P,Q,x}(t)}dt&
\end{eqnarray}
hold.

\begin{figure}[t]
 \begin{center}
 \hspace*{0em}\includegraphics[width=80mm, height=55mm]{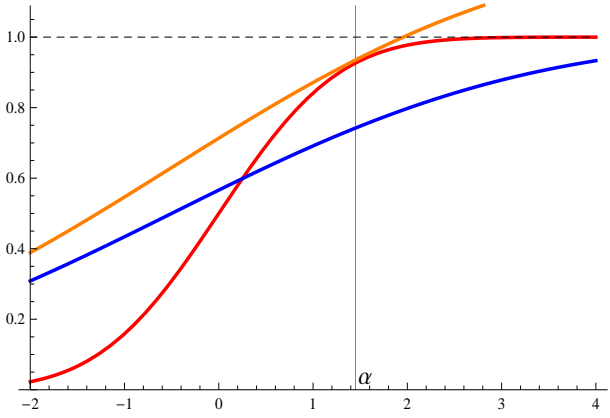}
 \end{center}
 \caption{
Let ${H(P)}/{V(P)}>{H(Q)}/{V(Q)}$. The red, the blue and the orange lines show $G_{P}$, $G_{P,Q,b}$ and $\frac{G_{P}(\alpha)}{G_{P,Q,b}(\alpha)}G_{P,Q,b}$, respectively. 
Then, $A_1$ is represented as the orange line on $x\le\alpha$ and the red line on $\alpha\le x$.
The limit of the maximal fidelity in Theorem \ref{ge1} coincides with the fidelity between  $A_1$ and the blue line $G_{P,Q,b}$.
}
 \label{Gauss1}
\end{figure}

We consider the approximation from $P^n$ to $Q^{ \frac{H(P)}{H(Q)}n+b\sqrt{n}}$.
There are three cases for the limit value of the maximal fidelity by the relation of  $\frac{H(P)}{V(P)}$ and $\frac{H(Q)}{V(Q)}$.
The following is the first case.
\begin{thm}\label{ge1}
When $\frac{H(P)}{V(P)}> \frac{H(Q)}{V(Q)}$, 
\begin{eqnarray}
\frac{N_{P}(x)}{N_{P,Q,b}(x)}=\frac{G_{P}(x)}{G_{P,Q,b}(x)}\label{threshold1}
\end{eqnarray}
has the unique solution $\alpha\in\R$ with respect to $x$, and the following holds.
\begin{eqnarray}
&&\hspace{-1em}{\lim}F^D(P^{ n}\to Q^{ \frac{H(P)}{H(Q)}n+b\sqrt{n}})\nonumber\\
&&\hspace{-2em}={\lim}F^M(P^{ n}\to Q^{ \frac{H(P)}{H(Q)}n+b\sqrt{n}})\\
&&\hspace{-2em}=\sqrt{G_{P}(\alpha)}\sqrt{G_{P,Q,b}(\alpha)}
+I_{P,Q,b}(\infty)-I_{P,Q,b}(\alpha)\\
&&\hspace{-2em}=:F_1(b)\label{F_1}
\end{eqnarray}
\end{thm}
For the continuous differentiable function
\begin{eqnarray}
&&A_1(x)
=\left\{
\begin{array}{ll}
\frac{G_{P}(\alpha)}{G_{P,Q,b}(\alpha)}G_{P,Q,b}(x) & \mathrm{if}~x\le \alpha \\
G_{P}(x) & \mathrm{if}~\alpha\le x,
\end{array}
\right.
\label{A_1}
\end{eqnarray}
the following equation holds
\begin{eqnarray}
F_1(b)
=F\left(\frac{dA_1}{dx},N_{P,Q,b}\right),\label{F_1 equation}
\end{eqnarray}
where $F\left(\frac{dA}{dx},N_{P,Q,b}\right)$ is the value defined as follows and is called the fidelity or the Bhattacharyya coefficient for continuous distributions.
\begin{eqnarray}
F\left(p,q\right)
:=\int_{\R}\sqrt{p(x)}\sqrt{q(x)}dx.
\end{eqnarray}
Therefore Theorem \ref{ge1} can be represented as 
\begin{eqnarray}
&&\hspace{-1.5em}{\lim}F^D(P^{ n}\to Q^{ \frac{H(P)}{H(Q)}n+b\sqrt{n}})
\nonumber\\
&&\hspace{-1.5em}={\lim}F^M(P^{ n}\to Q^{ \frac{H(P)}{H(Q)}n+b\sqrt{n}})
=F\left(\frac{dA_1}{dx},N_{P,Q,b}\right). \label{fidelity representation}
\end{eqnarray}
The positional relation of the functions $G_P, G_{P,Q,b}$ and $A_1$ is shown in Fig.\ref{Gauss1}.
This fact holds for the following theorems about the limit of the maximal fidelity.
In other words, the limit of the maximal fidelity can be represented by a continuous differentiable distribution function $A$ on $\R$ as follows. 
\begin{eqnarray}\label{AN}
&&\hspace{-2em}{\lim}F^D(P^{ n}\to Q^{ \frac{H(Q)}{H(P)}n+b\sqrt{n}})\nonumber\\
&&\hspace{-2em}={\lim}F^M(P^{ n}\to Q^{ \frac{H(Q)}{H(P)}n+b\sqrt{n}})
=F\left(\frac{dA}{dx},N_{P,Q,b}\right),
\end{eqnarray}
where a distribution function on $\R$ is defined as a right-continuous increasing function which satisfies $\displaystyle\lim_{x\to-\infty} A(x)=0$ and $\displaystyle\lim_{x\to\infty} A(x)=1$.

To prove Theorem \ref{ge1}, we give the sketch of proof of (\ref{fidelity representation}).
First, the following lemma is essential for the proof of the direct part. 
\begin{lem}\label{lem.direct}
Let $P$ and $Q$ be probability distributions. When a function $A$ on $\R$ is continuously differentiable, monotone increasing and $G_P\le A\le1$, the following holds. 
\begin{eqnarray}
{\rm liminf} F^D\left(P^n\to Q^{ \frac{H(Q)}{H(P)}n+b\sqrt{n}}\right)
\ge F\left(\frac{dA}{dx},N_{P,Q,b}\right).\label{direct}
\end{eqnarray}
\end{lem}
We do not provide the proof of the above lemma here.
By this lemma for $A=A_1$, the left term in (\ref{fidelity representation}) is greater than or equal to the right term.
Secondly, taking the limit superior in (\ref{fidelity ineq}), we obtain 
\begin{eqnarray}
{\limsup}F^D(P^{ n}\to Q^{ \frac{H(P)}{H(Q)}n+b\sqrt{n}})\nonumber\\
\le{\limsup}F^M(P^{ n}\to Q^{ \frac{H(P)}{H(Q)}n+b\sqrt{n}}).\label{lim fidelity ineq}
\end{eqnarray}
Therefore, the middle term in (\ref{fidelity representation}) is greater than or equal to the left term.
Lastly, we will explain that the right term in (\ref{fidelity representation}) is greater than or equal to the middle term.
We introduce the notation of subset in $\N$ as $\mathcal{A}_n(x):=\{1,2,\cdot\cdot\cdot,[ e^{H(P)n+x\sqrt{n}}]\}$ and $\mathcal{A}_n(x,x'):=\mathcal{A}_n(x')\setminus\mathcal{A}_n(x)$.
Then, for an arbitrary $0<\epsilon$ and probability distributions $\{P'_n\}_{n=1}^{\infty}$ such that $P_n$ is majorized by each $P'_n$, let us take a real number $c\in\R$ which satisfies 
\begin{eqnarray}
&&\sqrt{1-G_{P}(c)}\sqrt{1-G_{P,Q,b}(c)}<\epsilon.
\end{eqnarray}
and $\alpha<c$ where $\alpha$ is the unique solution in (\ref{threshold1}).
Moreover, for a natural number $I$, let $x_i^I$ be $\alpha+\frac{c-\alpha}{I}i$.
Since the monotonicity of the fidelity, 
\begin{eqnarray}
&&F(P'_n,Q_n)\nonumber\\
&&\hspace{-1em}\le\sqrt{P'_n(\mathcal{A}_n(x_{0}))} \sqrt{Q_n(\mathcal{A}_n(x_{0}))}\nonumber\\
&&+\sum_{i=1}^{I}\sqrt{P'_n(\mathcal{A}_n(x_{i-1}^I,x_{i}^I))} \sqrt{Q_n(\mathcal{A}_n(x_{i-1}^I,x_{i}^I))}\nonumber\\
&&+\sqrt{P'_n(\mathcal{A}_n(c,\infty))} \sqrt{Q_n(\mathcal{A}_n(c,\infty))}.\label{monotonicity}
\end{eqnarray}
By direct calculation, when $n$ goes to $\infty$
\begin{eqnarray}
&&\hspace{-1em}\mathrm{limsup}F(P^n\to Q^{ \frac{H(Q)}{H(P)}n+b\sqrt{n}})\nonumber\\
&&\hspace{-2em}\le \sqrt{G_{P}(\alpha)}\sqrt{G_{P,Q,b}(\alpha)}+\int_{\alpha}^{c}\sqrt{N_{P}(x)}\sqrt{N_{P,Q,b}(x)}dx\label{division}\\
&&\hspace{-1em}+\sqrt{1-G_{P}(c)}\sqrt{1-G_{P,Q,b}(c)}\nonumber\\
&&\hspace{-2em}\le \sqrt{G_{P}(\alpha)}\sqrt{G_{P,Q,b}(\alpha)}+\int_{\alpha}^{\infty}\sqrt{N_{P}(x)}\sqrt{N_{P,Q,b}(x)}dx+\epsilon\nonumber\\
&&\hspace{-2em}= F\left(\frac{dA_1}{dx},N_{P,Q,b}\right)+\epsilon \label{converse}
\end{eqnarray}
holds.
Therefore, the right term in (\ref{fidelity representation}) is greater than or equal to the middle term.
Taken together, we obtain (\ref{fidelity representation}) from (\ref{direct}), (\ref{lim fidelity ineq}) and (\ref{converse}).

\begin{thm}\label{ge1 rate}
When $\frac{H(P)}{V(P)}> \frac{H(Q)}{V(Q)}$, the second order asymptotic expansion for a confidence coefficient $0<\nu<1$ in (\ref{asym.expansion2}) is described as follows. 
\begin{eqnarray}\label{exp.ge1}
&\hspace{-1em}L^D_n(P,Q|\nu)
=(H(P)/H(Q))n-F_1^{-1}(\nu)\sqrt{n}+o(\sqrt{n}).&\nonumber
\end{eqnarray} 
In particular, $R_2(P,Q|\nu)=R^D_2(P,Q|\nu)=F_1^{-1}(\nu)$.
\end{thm}

\begin{figure}[t]
 \begin{center}
 \hspace*{0em}\includegraphics[width=80mm, height=55mm]{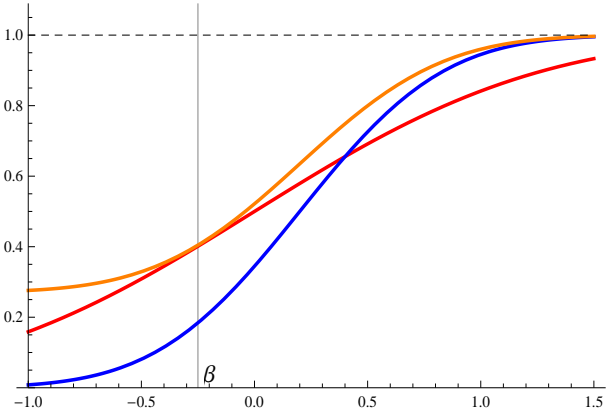}
 \end{center}
 \caption{
Let ${H(P)}/{V(P)}<{H(Q)}/{V(Q)}$. The red, the blue and the orange lines show $G_{P}$, $G_{P,Q,b}$ and $1-\frac{1-G_P(\beta)}{1-G_{P,Q,b}(\beta)}(1-G_{P,Q,b}(x))$, respectively. 
Then, $A_2$ is represented as the red line on $x\le\beta$ and the orange line on $\beta\le x$.
The limit of the maximal fidelity in Theorem \ref{le1} coincides with the fidelity between  $A_2$ and the blue line $G_{P,Q,b}$.
}
 \label{Gauss2}
\end{figure}

The following treat the second case on the limit value of the maximal fidelity.
\begin{thm}\label{le1}
When $\frac{H(P)}{V(P)}< \frac{H(Q)}{V(Q)}$, 
\begin{eqnarray}\label{threshold2}
\frac{N_{P}(x)}{N_{P,Q,b}(x)}=\frac{1-G_{P}(x)}{1-G_{P,Q,b}(x)}
\end{eqnarray}
has the unique solution $\beta\in\R$ with respect to $x$, and the following holds.
\begin{eqnarray}
&&\hspace{-1em}{\lim}F^D(P^{ n}\to Q^{ \frac{H(Q)}{H(P)}n+b\sqrt{n}})\nonumber\\
&&\hspace{-2em}={\lim}F^M(P^{ n}\to Q^{ \frac{H(Q)}{H(P)}n+b\sqrt{n}})\nonumber\\
&&\hspace{-2em}=I_{P,Q,b}(\beta)
+\sqrt{1-G_{P}(\beta)}\sqrt{1-G_{P,Q,b}(\beta)}\nonumber\\
&&\hspace{-2em}=:F_2(b)\label{F_2}
\end{eqnarray}
\end{thm}
Taking the following function as a function $A:\R\to[0,1]$, Theorem \ref{le1} can be represented as (\ref{AN}). 
\begin{eqnarray}
&A_2(x)
=\left\{
\begin{array}{ll}
G_{P}(x) & \mathrm{if}~x\le \beta \\
1-\frac{1-G_P(\beta)}{1-G_{P,Q,b}(\beta)}(1-G_{P,Q,b}(x)) & \mathrm{if}~\beta\le x .
\end{array}
\right.&
\label{A_2}\nonumber
\end{eqnarray}
The positional relation of $G_P, G_{P,Q,b}$ and $A_2$ is shown in Fig.\ref{Gauss2}. 
Theorem \ref{le1} can be proven as the same as Theorem \ref{ge1} by the combination of (\ref{direct}), (\ref{lim fidelity ineq}) and (\ref{converse}).
But, we need to correct the threshold points $\alpha$ and $c$ in (\ref{converse}) and change the area where the function $\sqrt{N_P}\sqrt{N_{P,Q,x}}$ is integrated.

\begin{thm}\label{le1 rate}
When $\frac{H(P)}{V(P)}< \frac{H(Q)}{V(Q)}$,
the second order asymptotic expansions for a confidence coefficient $0<\nu<1$ in (\ref{asym.expansion2}) are described as follows. 
\begin{eqnarray}\label{exp.le1}
&L^D_n(P,Q|\nu)
=(H(P)/H(Q))n-F_2^{-1}(\nu)\sqrt{n}+o(\sqrt{n}).&\nonumber
\end{eqnarray} 
In particular, $R_2(P,Q|\nu)=R^D_2(P,Q|\nu)=F_2^{-1}(\nu)$.
\end{thm}

\begin{figure}[t]
 \begin{center}
 \hspace*{0em}\includegraphics[width=80mm, height=55mm]{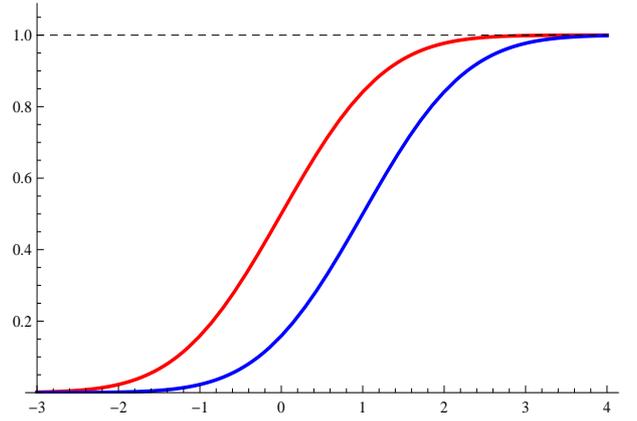}
 \end{center}
 \caption{
Let ${H(P)}/{V(P)}= {H(Q)}/{V(Q)}$. The red and blue lines show $G_{P}$ and $G_{P,Q,b}$, respectively. The limit of the maximal fidelity in Theorem \ref{=1} coincides with the fidelity between the red line $G_{P}$ and the blue line $G_{P,Q,b}$
}
 \label{Gauss3}
\end{figure}

The following treat the third case on the limit value of the maximal fidelity.
\begin{thm}\label{=1}
When $\frac{H(P)}{V(P)}=\frac{H(Q)}{V(Q)}$, the following holds.
\begin{eqnarray}
\hspace{-1em}&&{\lim}F^D(P^{ n}\to Q^{ \frac{H(Q)}{H(P)}n+b\sqrt{n}})\nonumber\\
\hspace{-1em}&&={\lim}F^M(P^{ n}\to Q^{ \frac{H(Q)}{H(P)}n+b\sqrt{n}})
=\left\{
\begin{array}{l}
 e^{\frac{-(H(Q)b)^2}{8V(P)}}~ \mathrm{if}~b<0 \\
1 \hspace{4em}\mathrm{if}~b\ge0
\end{array}
\right.\nonumber
\end{eqnarray}
\end{thm}
Taking the cumulative distribution function $G_P$ as a function $A:\R\to[0,1]$Theorem \ref{=1}can be represented as (\ref{AN}). 
The positional relation of $G_P, G_{P,Q,b}$ is shown in Fig.\ref{Gauss3}. 
Theorem \ref{=1} can be proven as the same as Theorem \ref{ge1} by the combination of (\ref{direct}), (\ref{lim fidelity ineq}) and (\ref{converse}).
But, we need to change the area where the function $\sqrt{N_P}\sqrt{N_{P,Q,x}}$ is integrated to the whole of real numbers.

\begin{thm}\label{=1 rate}
When $\frac{H(P)}{V(P)}= \frac{H(Q)}{V(Q)}$,
the second order asymptotic expansions for a confidence coefficient $0<\nu<1$ in (\ref{asym.expansion2}) are described as follows. 
\begin{eqnarray}\label{exp.=1}
& \hspace{-1em}L^D_n(P,Q|\nu)
=(H(P)/H(Q))n-\sqrt{\frac{8V(P)\mathrm{log}\nu^{-1}}{H(Q)}}\sqrt{n}+o(\sqrt{n}).&\nonumber
\end{eqnarray} 
In particular,
\begin{eqnarray}
R_2(P,Q|\nu)=R^D_2(P,Q|\nu)=\sqrt{\frac{8V(P)\mathrm{log}\nu^{-1}}{H(Q)}}.
\end{eqnarray}
\end{thm}

We note that Theorems \ref{ge1} and \ref{le1} which we proved coincide with Theorem \ref{2-order error2} and Theorem \ref{2-order error} in the limit $Q\to U_2$ and $P\to U_2$, respectively.
Therefore, our results can be regard as extensions of existing studies in the limit.

\section{Conclusion}

We treated the second order asymptotics of the random number generation from an i.i.d. probability distribution of P to that of $Q$. 
In existing studies, $P$ or $Q$ has been assumed to be a uniform distribution, but in this paper, both probability distributions have not been restricted to a uniform distribution.
Let us review the proof of Theorem \ref{ge1} in which we derived the limit of the maximal fidelity.
Other theorems after Theorem \ref{ge1} were variants of or derived from Theorem \ref{ge1}.
In the direct part, the maximal fidelity was achievable by performing different transformation to $P^n$ on each area which was divided at points $\alpha$ and $c$.
But we emphasize that (\ref{division}) is actually valid for any real numbers $\alpha$ and $c$.
Therefore, the choice of $\alpha$ and $c$ is not important in the direct part but essential in the direct part. 
In addition, the notion of the majorization was the key of the proof of the converse part
since transformations in the sense of the majorization was wider than deterministic transformations, and thus the maximal fidelity under the majorization condition gave an upper bound the maximal fidelity under deterministic transformations as (\ref{fidelity ineq}).  
Finally, we  remark that the majorization relation has an operational meaning as a transformation called LOCC for quantum entangled states in the quantum information theory, and our results can be extended to the quantum settings.
A part of extensions of our results to quantum information theory is treated in \cite{BBPS,Hay2, JP}.

\section*{Acknowledgment}

 WK acknowledges support from Grant-in-Aid for JSPS Fellows No. 233283. MH is partially supported by a MEXT Grant-in-Aid for Scientific Research (A) No. 23246071. 
The Center for Quantum Technologies is funded by the Singapore Ministry of Education and the National Research Foundation as part of the Research Centres of Excellence programme.

\end{document}